\documentstyle[aps,prd,epsfig]{revtex}
\newcommand{\beq}{\begin{equation}}
\newcommand{\eeq}{\end{equation}}
\newcommand{\be}{\begin{equation}}
\newcommand{\ee}{\end{equation}}
\newcommand{\bea}{\begin{eqnarray}}
\newcommand{\eea}{\end{eqnarray}}
\newcommand{\ba}{\begin{array}{ccc}}
\newcommand{\ea}{\end{array}}
\newcommand{\nn}{\nonumber}
\newcommand{\sigvev}{\langle \sigma \rangle}
\newcommand{\Svev}{\langle D \rangle}
\newcommand{\pimvev}{\langle \pi \rangle}

\def\beqn{\begin{eqnarray}}
\def\eeqn{\end{eqnarray}}

\def\al{\varphi}
\def\d{\partial}
\def\Tr{ {\rm Tr} }

\begin{document}

\title{The Superfluid and Conformal Phase Transitions of Two-Color QCD}
\vfill
\author{J.~T. {\sc Lenaghan}$^{(a)}$\footnote{Electronic address:
{\tt lenaghan@alf.nbi.dk} } \quad F. {\sc Sannino}$^{(b)}$
\footnote{Electronic address: {\tt sannino@alf.nbi.dk}} \quad K.
{\sc Splittorff}$^{(a)}$ \footnote{Electronic address: {\tt
split@alf.nbi.dk}}} \vfill \address{~\\ {\rm $^{(a)}$ The Niels
Bohr Institute \& $^{(b)}$NORDITA} \\Blegdamsvej 17, DK-2100
Copenhagen \O, Denmark}
\date{July 2001}
\maketitle

\begin{abstract}
The phase structure of two-color QCD is examined as a function of the chemical
potential and the number of light quark flavors.  We consider
effective Lagrangians for two-color QCD containing the Goldstone
excitations, spin-one particles and negative intrinsic parity terms.
We discuss the possibility of a conformal phase transition and the
enhancement of the global symmetries as the number of flavors is
increased. The effects of a quark chemical potential on the spin-one
particles  and on the negative intrinsic parity terms are analyzed.
It is shown that the phase diagram that is predicted by
the linearly realized effective Lagrangian at tree-level matches
exactly that predicted by chiral perturbation theory.
\end{abstract}

\section{Introduction}

Quantum chromodynamics (QCD) at large quark chemical potential has
attracted a great deal of interest in recent years \cite{REV}.
Since single gluon exchange between two quarks is attractive in
the color-antitriplet channel \cite{Barrois:1977xd,Bailin:1984bm},
quark matter
is expected to behave as a color superconductor for a sufficiently
large quark chemical potential.  Possible phenomenological
applications include the description of quark stars, neutron star
interiors and the physics near the core of collapsing stars
\cite{REV,OS,HHS}. {}From a theoretical point of view, one would
like to be able to derive the QCD phase diagram from first
principles as a function of temperature, chemical potential and
the number of light flavors.  While much has been learned about
the phase structure of QCD at nonzero temperature through a
combination of perturbation theory and lattice simulations, the
phase structure at nonzero chemical potential and for large
numbers of flavors has been less extensively explored
\cite{MariapaolaI}.  The phase structure as the number of light
flavors is increased is expected to be quite rich. {}For example,
in Ref.\ \cite{mawhinney}, the effects of chiral symmetry breaking
were found to be dramatically reduced as the number of flavors was
increased from zero to four. At asymptotically large quark
chemical potentials, $\mu\gg\Lambda_{\rm QCD}$, perturbation
theory is valid and one is able to perform controlled
calculations.  {}For small to intermediate chemical potentials,
however, one must rely either on effective theories or perform
lattice simulations. Standard importance sampling methods employed
in lattice simulations fail, however, at nonzero chemical
potential for $N_c=3$ since the fermionic determinant is complex.

{}For $N_c=2$, the situation is very different since the quarks
are in a pseudoreal representation of the gauge group.  The first
difference is that the fermionic determinant in the path integral
is real (thought not necessarily positive) and so lattice
simulations can be performed at nonzero baryon and isospin
chemical potential for an even number of quark flavors
\cite{Morrison:1998ud,Hands:1999md,Aloisio:2000nr,Liu:2000in,Aloisio:2000if,Bittner:2001rf,Hands:2001hi,Muroya:2001qp,Hands:2001yh,Aloisio:2001rb,Kogut:2001if,Kogut:2001na}.  Secondly, $N_f^2-N_f$ of the Goldstone
excitations are diquarks which carry nonzero baryon charge.  As
shown in Refs.\ \cite{KST,KSTVZ}, this has the advantage that
chiral perturbation theory \cite{GL} is valid at the critical
chemical potential as opposed to the case of $N_c=3$ for which the
critical chemical potential lies well above the scale at which
chiral perturbation theory becomes invalid.  Also, the formation
of a diquark condensate in the $N_c=2$ theory does not break gauge
invariance and so exhibits superfluidity at large chemical
potentials, unlike the case of $N_c=3$ which exhibits
superconductivity.

There also has been much progress in understanding the phase structure of
supersymmetric theories as the number of massless fermions is varied
\cite{IS}.  While much less in known for nonsupersymmetric theories,
the infrared behavior of such theories should change dramatically as
the number of massless fermions is increased.  In particular, for $N_f
> 11$, the one-loop beta function for $N_c=2$ QCD becomes positive and
the theory loses asymptotic freedom.  In this non-Abelian QED-like
phase, the theory is not expected to be confining or to exhibit
chiral symmetry breaking.  Just below $N_f=11$, a perturbative
infrared stable fixed point develops.  In this phase, the trace of
the energy momentum tensor vanishes and the theory is a
non-Abelian conformal field theory.  Just before the onset of the
conformal phase, it has been argued that an enhanced global
symmetry can emerge involving the massive spectrum of the theory
\cite{ARS,AS}. This new dynamical symmetry may arise when the
number of light flavors is near a critical number of flavors
(about $N_F=$8 \cite{ARS,AS,ATW,FSS}).  Above this value, the
theory is expected to enter the conformal regime.  This enhanced
symmetry group may be important, for example, when coupling a
two-color strongly interacting theory to the electroweak symmetry
breaking sector of the standard model \cite{ARS}.  The hadronic
spectrum close to this point can be very different than for a
smaller number of flavors.  {}For example, this enhanced global
symmetry can lead to degenerate masses for the vector spectrum
even in the presence of chiral symmetry breaking.  Additionally,
when approaching the conformal point all of the massive states of
the theory become light exponentially fast as the number of
flavors reaches the critical value.

In this work, we study the phase structure of $N_c=2$ QCD as a
function of chemical potential and the number of light flavors.  We
begin by reviewing the most general effective Lagrangian for $N_c=2$
describing the Goldstone bosons and spin-one states and their
interactions.  This Lagrangian was first constructed in Refs.\
\cite{ARS} and \cite{DRS} within the context of extended technicolor
theories.  We first discuss the nonlinear realization of chiral
symmetry and later consider the linear effective Lagrangian.  Before
introducing a chemical potential, the possible phase structure of
two-color QCD as the number of flavors is increased is discussed.  The enhanced
global symmetry mentioned above is identified at the level of the
effective Lagrangian.

We stress that if we were to consider an approximate local flavor
symmetry when introducing the spin-one particles, then no enhanced
symmetry is allowed. An intriguing possibility is that for a small
number of flavors relative to the number of colors it might be
reasonable to include the vectors as almost gauge vectors of
chiral symmetry, while for a large number of flavors the enhanced
global symmetry may set in. Both limits severely constrain the
effective Lagrangians. Lattice simulations are a very useful means
of testing such conjectures.


An essential component for any effective Lagrangian for a strongly
interacting theory is the set of intrinsic negative parity terms,
i.e. those terms contracted by the fully antisymmetric tensor
$\epsilon_{\eta\nu\rho\sigma}$.  The Wess--Zumino term
\cite{WZ,Witten} is the time honored example of an $\epsilon$ term
and is needed to saturate (in the Goldstone phase) the 't Hooft
global non-Abelian anomaly constraints.  When the underlying
fermions are in a pseudoreal representation of the gauge group,
the $\epsilon$ part of the effective Lagrangian involving the
Goldstone bosons and the spin-one fields was explicitly
constructed in Ref.\ \cite{DRS}.  We show that this sector of the
theory couples to the baryon and isospin chemical potentials.
These terms are also expected to be important when studying the
solitonic sector of the theory.

Next, we study the effect of the chemical potential on the
spin-one fields and we explicitly calculate the mass gaps.  The
results suggest that some of the vectors may condense and hence
break rotational invariance. The value of the chemical potential
at the onset of the vector condensation is predicted by assuming
that vectors are sufficiently massive and that the vacuum
alignment of the theory is determined by the Goldstone
excitations.

We then turn to study the superfluid phase transition at nonzero
chemical potential using a linear realization of the underlying chiral
symmetry.  The linear sigma model not only has the advantage that the
calculations are relatively simple but also allows us to predict how
the magnitude of the condensates changes with the baryon and isospin
chemical potentials. The vacuum structure that is predicted at
tree-level is identical to that of chiral perturbation theory
\cite{Splittorff:2001mm}.
The dispersion relations of the pseudo-Goldstone bosons
are calculated as well as the dependence of the condensates on the
baryon and isospin chemical potentials.

This paper is organized as follows.  In Sec.\ \ref{sec:nonlinrev},
the nonlinearly realized effective Lagrangian is reviewed. In
Sec.\ \ref{sec:conformal} we briefly comment on the $N_f$ phase
diagram at zero chemical potential and review the possibility
\cite{ARS,AS,DRS} that novel phenomena like parity doubling can
emerge near the conformal phase transition.  The effects of a
nonzero baryon chemical potential are illustrated in Secs.\
\ref{sec:baryonden}, \ref{sec:e-terms-mu}, and
\ref{sec:vec-and-mu}. {}First, we study the effects of the
chemical potential on the $\epsilon$ terms and then we consider
the dispersion relations of the spin-one fields in the presence of
a chemical potential. We then turn to the linearly realized
theory. In Sec.\ \ref{sec:linrev}, the linear effective Lagrangian
is reviewed.  The phase diagram and dispersion relations at
nonzero baryon and isospin chemical potentials are studied and the
results are contrasted with those of chiral perturbation theory.
We conclude in Sec.\ \ref{sec:conc}. Our conventions are
summarized in the Appendix.

\section{The Two-color Nonlinear Effective Lagrangian}
\label{sec:nonlinrev}

The simplest example of a gauge theory with fermions in a
pseudoreal representation is $N_c=2$ QCD with the fermions in the
fundamental representation.  The quantum global symmetry for $N_f$
matter fields is $SU(2N_f)$ which contains $SU_L(N_f)\times
SU_R(N_f)$. Using the Wess and Bagger spinorial conventions
\cite{WB}, the underlying Lagrangian is
\begin{equation}
{\cal L}_{N_c=2}=-\frac{1}{4g^2} \vec{G}_{\mu\nu}\cdot\vec{G}^{\mu\nu} +
i\, \bar{Q} \bar{\sigma}^{\nu}\left[\partial_{\nu} - i\,
\vec{G_{\nu}}\cdot \frac{\vec{\tau}}{2} \right] Q -
\frac{1}{2}m_{q}Q^T \tau_2 \, E \, Q  + {\rm h.c.} \ .
\label{Lqcd}
\end{equation}
$G_{\mu\nu}^a$ and $G_{\nu}^a$ with $a=1,2,3$ are the
gluon field strength and field respectively, while the $\tau^a$ are the Pauli
matrices for the $SU_c(2)$ group. $Q^{c,I}_{\alpha}$ is a
two-spinor fermion field in the fundamental representation of
color with $c=1,2$ and $I=1,\cdots, 2N_f$,
\be
 Q =\left(\begin{array}{c}
 q_L \\ i\sigma_2\tau_2 q^*_R
\end{array}\right) \ .
\ee
In the massless limit ($m_q=0$), the classical global symmetry is
$U(2N_f)$ which is then broken by the Adler-Bell-Jackiw anomaly to
$SU(2N_f)$. The mass term explicitly breaks the $SU(2N_f)$ symmetry to
$Sp(2N_f)$.  The $2N_f\times 2N_f$ matrix $E$ is
\begin{equation}
E=\left(
\begin{array}{cc}
{\bf 0} & {\bf 1} \\ -{\bf 1} & {\bf 0} \label{E}
\end{array}
\right) \ .
\end{equation}
{}For a sufficiently small number of flavors, one expects the theory
to confine and to dynamically generate a condensate which
spontaneously breaks the global symmetry group for zero quark masses.
For three-color QCD with quarks in the fundamental representation,
this condensate is the usual quark-antiquark condensate; however, on
account of the enlarged $SU(2N_f)$ global symmetry in two-color QCD,
any quark--antiquark condensate can be continuously rotated into a
quark--quark condensate and the only discernible condensate is an
admixture of these two condensates.  The subgroup down to which the
$SU(2N_f)$ symmetry is broken is usually taken to be the maximal
diagonal subgroup $Sp(2 N_f)$ \cite{Peskin}, a choice which is
consistent with a new criterion presented in Ref.~\cite{ADS}.

We now turn to the construction of the low energy effective
Lagrangian.  We divide the hermitian generators, $\{T^a\}$, of
$SU(2N_f)$, normalized according to $\displaystyle{{\rm
Tr}\left[T^aT^b\right]}=\delta^{ab}/2$, into two classes: the
generators of $Sp(2N_f)$ which we denote by $\{S^{a}\}$ with
$a=1,\ldots ,2N_f^2+N_f$, and the remaining generators of
$SU(2N_{f})$ which we denote by $\{X^{i}\}$ with $i=1,\ldots
,2N_f^2-N_f-1$.  Note that the latter set parameterizes the
quotient space $SU(2N_{f})/Sp(2N_{f})$.  An explicit realization
of the generators is provided in the Appendix.  This breaking pattern
gives $2N^2_f - N_f -1$ Goldstone bosons which are encoded in the
$2N_f\times2N_f$ antisymmetric matrix
\begin{equation}
U=e^{i\frac{\Pi ^{j}X^{j}}{v}}\,E\ .
\label{parametrizationI}
\end{equation}
$U$ transforms linearly under a chiral rotation as
\begin{equation}
U\rightarrow u\, U \, u^{T} \ ,
\end{equation}
with $u\in SU(2N_f)$. The nonlinear realization constraint,
$\displaystyle {UU^{\dagger }=1}$, is automatically satisfied.

The generators of the $Sp(2N_f)$ satisfy the relation
\begin{equation}
S^{T}\,E+E\,S=0\ ,
\label{SE}
\end{equation}
while the $X^i$ generators obey
\begin{equation}
X^{T}\,E-E\,X=0\ . \label{XE}
\end{equation}
Using this last relation we can easily demonstrate that
$U^{T}=-U$. {}For simplicity, we also require that
\begin{equation}
{\rm Pf}\,U=1\ ,
\end{equation}
in order to avoid discussing the explicit realization of the
underlying Adler-Bell-Jackiw axial anomaly at the effective
Lagrangian level \cite{FSS}.

\subsection{The Spin-One Fields}

We next introduce the coupling between the Goldstone excitations and a
vector field.  While there are many different ways to introduce vector
fields at the level of the effective Lagrangian (the hidden local
gauge symmetry of Ref.\ \cite{BKY}, for example), they are all
equivalent at tree-level.  We consider the vector field
\begin{equation}
A_{\nu }=A_{\nu }^{a}T^{a}\ ,
\label{A}
\end{equation}
which we take to transform under a $SU(2N_{f})$ rotation as
\begin{equation}
A_{\nu }\rightarrow uA_{\nu }u^{\dagger }-i(\partial _{\nu
}u)u^{\dagger }\ .
\label{transf-A}
\end{equation}
It is useful to formally define a chiral covariant derivative
\begin{equation}
D_{\nu }U=\partial _{\nu }U-iA_{\nu }U-iUA_{\nu }^{T}\ .
\end{equation}
The most general two-derivative term which preserves
local chiral symmetry is
\begin{equation}
{\rm Tr}\left[ D_{\nu}UD^{\nu}U^{\dagger
}\right] \ .
\end{equation}
Although we have introduced the vector fields formally as vectors
associated with a local chiral gauge theory, the effective
Lagrangian must respect the global $SU(2N_{f})$
transformations and is given by
\begin{eqnarray}
{\cal L}_{eff}&=&v^{2}\,{\rm Tr}\left[ D_{\nu }UD^{\nu }U^{\dagger }\right]
+m_V^{2}\, {\rm Tr}\left[ A_{\nu }A^{\nu }\right]
+ hv^{2}\,{\rm Tr}\left[ A_{\nu }UA^{T\nu }U^{\dagger }\right] \nonumber \\
&+& i\,sv^{2}\, {\rm Tr}\left[ A_{\nu }UD^{\nu }U^{\dagger }\right]
+v^2m_\pi^2\Tr[{\cal
  M}U+{\cal M}^\dagger U^\dagger]
\ .  \label{nr2}
\end{eqnarray}
We counted the vector fields as derivatives and added a democratic
quark mass matrix
\beqn \label{eq:Mmatrix} {\cal M}
\equiv \left(\begin{array}{cc} {\bf 0} &{\bf - 1}
\\ {\bf 1} & {\bf 0} \end{array}\right)  \ .
\eeqn The parameters $m_V$, $s$, $h$ are real constants which
effectively measure the departure from local chiral symmetry.
The degenerate masses of the pseudo-Goldstone exicitations
are denoted by $m_\pi$.  
Equation (\ref{nr2}) is the most general
two-derivative effective Lagrangian compatible with the global
symmetries of $N_c=2$ QCD \cite{ARS,DRS}.

{}For completeness, we augment the effective Lagrangian with the
simplest possible kinetic term for the vectors:
\begin{mathletters}
\begin{equation}\label{eq:kin}
{\cal L}_{kin}=-\frac{1}{2g^{2}}{\rm Tr}\left[ F_{\rho \nu
}F^{\rho \nu } \right] \ ,
\end{equation}
where
\begin{equation}\label{eq:F}
F^{\rho \nu }=\partial ^{\rho }A^{\nu }-\partial ^{\nu }A^{\rho
}-i\left[ A^{\rho },A^{\nu }\right] \ .
\end{equation}
\end{mathletters}
The vector kinetic piece arises as a fourth order term in the
derivative counting and $g$ is a dimensionless coupling constant.
The tree-level masses of the vectors can also
be calculated and are given by
\begin{mathletters}
\bea
M_{S}^{2} &=& g^2(m_V^2 - h v^2) \\
M_{X}^{2} &=& g^2(m_V^2 +v^2 ( h + 4 -2 s)) \,\, ,
\eea
\end{mathletters}
where we have split the vectors into those associated
with the $\{S^a\}$ generators and those associated with
the $\{X^a\}$ generators.  Note that, in general, there
is a nonzero mass splitting between these two sets.

\subsection{The ${\protect\epsilon}$ terms for $SU(2N_{f})$}

Next, we consider the complete set of 4-derivative terms which
have negative intrinsic parity and contain spin-one and spin-zero
fields. As mentioned in the introduction, these terms contain the
Lorentz antisymmetric tensor $\epsilon_{\eta \nu \rho \sigma}$ and
the canonical example is the Wess-Zumino action.  These terms are
necessary at the effective Lagrangian level since they account for
the 't Hooft global anomaly constraints in the Goldstone phase.
Additionally, they are important when quantizing the solitonic
sector of the theory.  They can be compactly written using the
language of algebra-valued differential forms:
\begin{equation}
\alpha =\left( dU\right) U^{-1}=\left(\partial_{\nu}U \right) U
^{-1}dx^{\nu}\ .
\end{equation}
Since the fermions are in a pseudoreal representation of the gauge
group, it is sufficient to define only one type
of differential form, $\alpha$, since now the other possible form
$\beta =U^{-1}dU=\alpha ^{T}$ is not independent \cite{DRS}. The
Wess-Zumino term is
\begin{equation}
{\Gamma}_{WZ}\left[ U\right] =C\,\int_{M^{5}}{\rm Tr}\left[ \alpha
^{5} \right] \ .  \label{WZSp}
\end{equation}
The dimension of the spacetime must be increased by one spatial
direction in order to make the action local.  Hence, the integral
in eq.\ (\ref{WZSp}) must be performed over a five-dimensional
manifold whose boundary ($M^4$) is the ordinary four-dimensional
Minkowski space. The coefficient $C$ can be fixed, in general, by
matching the anomalous variation of the currents associated with
non-Abelian anomalies at the effective Lagrangian level.  {}For
the case at hand, the constant $C$ is fixed to be
\begin{equation} \label{eq:Cval}
C=-i\frac{N_c}{240\pi^2} \ ,
\end{equation}
with $N_c=2$ \cite{DRS}. We note that the coefficient is similar to
the case $N_c=3$ since $SU(2N_f)\supset SU_L(N_f)\times SU_R(N_f)$.

Since we are considering a theory which contains vectors and only
global chiral symmetry, eq.\ (\ref{WZSp}) needs to be generalized.
This has been done in Ref.\ \cite{DRS}. In order to generate all
allowed terms, the authors of Ref.\ \cite{DRS} first formally gauged
the Wess--Zumino action following a standard procedure developed in
Refs.\ \cite{Witten,KRS,KS}.  This procedure automatically provides
most of the desired terms and local chiral invariance relates
the coefficients of the new $\epsilon$ terms to the Wess--Zumino
coefficient. The effective Lagrangian was then generalized to be only
globally invariant under chiral rotations, and as a result, all the
terms have different coefficients. Remarkably, the gauging procedure
generates all but one term allowed by global invariance.

The most general 4-derivative $\epsilon$ Lagrangian respecting
global chiral rotations is
\begin{eqnarray}
{\Gamma}_{WZ}\left[ U,\;A\right] &=&{\Gamma}_{WZ}\left[ U\right]
\,+i 10\,C_{1}\int_{M^{4}}{\rm Tr}\left[ A\alpha ^{3}\right]
\nonumber \\ &&-10\,C_{2}\,\int_{M^{4}}{\rm Tr}\left[
(dAA+AdA)\alpha \right] \nonumber \\ &&-5\,C_{3}\int_{M^{4}}{\rm
Tr}\left[ dAdUA^{T}U^{-1}-dA^{T}dU^{-1}AU \right] \nonumber \\
&&-5\,C_{4}\,\int_{M^{4}}{\rm Tr}[UA^{T}U^{-1}(A\alpha ^{2}+\alpha
^{2}A)] \nonumber \\ &&+5\,C_{5}\int_{M^{4}}{\rm Tr}\left[
(A\alpha )^{2}\right] +i 10\,C_{6}\,\int_{M^{4}}{\rm Tr}\left[
A^{3}\alpha \right]  \nonumber \\ &&+i 10\,C_{7}\,\int_{M^{4}}{\rm
Tr}\left[ (dAA+AdA)UA^{T}U^{-1}\right]  \nonumber
\\
&&-i 10\,C_{8}\int_{M^{4}}{\rm Tr}\left[ A\alpha
AUA^{T}U^{-1}\right] +10\,C_{9}\,\int_{M^{4}}{\rm
Tr}[A^{3}UA^{T}U^{-1}]  \nonumber \\
&&+\frac{5}{2}C_{10}\int_{M^{4}}{\rm Tr}[(AUA^{T}U^{-1})^{2}]
\nonumber
\\ &&+iC_{11}\int_{M^{4}}{\rm Tr}[A^{2}(\alpha
UA^{T}U^{-1}-UA^{T}U^{-1}\alpha )] \ , \label{generalWZ}
\end{eqnarray}
where $C_{i}$ are imaginary coefficients and $A=A^{\nu}dx_{\nu}$
\cite{DRS}.  It is important to stress that when imposing local
chiral symmetry all of the coefficients are given by eq.\
(\ref{eq:Cval}) except for $C_{11}$ which is zero\footnote{NB: We
have changed the normalization of the coefficients with respect to
Ref.\cite{DRS}}. Using C and CP invariance, one can show that
there are no other negative intrinsic odd parity terms at this
order. Aside from the standard four-derivative terms involving the
Goldstone fields, we are now endowed with a rather complete and
general effective Lagrangian.  The action is
\be
S_{eff} = \int d^4x \left( {\cal L}_{eff} + {\cal L}_{kin}\right)
    + \Gamma_{WZ}[U,A] \ .
\ee

\section{The Phase Structure Along the $N_f$-axis}
\label{sec:conformal}

As mentioned in the Introduction, the infrared behavior of gauge
theories changes dramatically as the number of light fermion
flavors is varied \cite{IS}.  In this section, we review two
possibilities that should be directly accessible in lattice
simulations of two-color QCD. Predictions resulting from analytic
calculations are at the moment only possible for supersymmetric
theories; however, the behavior of the beta function for two-color
QCD does offer some guidance.  We focus here on zero chemical
potential, but in later sections will briefly discuss the
large-$N_f$ behavior of two-color QCD at nonzero chemical
potential.

\subsection{The Conformal Phase Transition}

\begin{figure}[htb]
\begin{center}
\leavevmode {\epsfxsize=15.0truecm \epsfbox{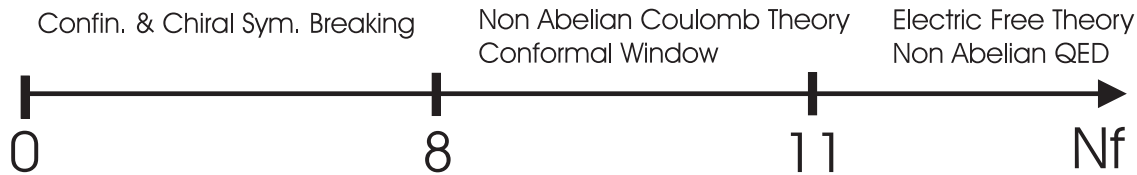}}
\end{center}
\caption{The possible phase structure of $N_c=2$ QCD as function of the number of
light quark flavors.}\label{fig1}
\end{figure}

{}For $N_f > 11$, the one-loop beta function of two-color QCD
changes sign and the theory loses asymptotic freedom
\cite{Gross:1973id,Politzer:1973fx}.  The resulting infrared free
theory is now in a non-Abelian QED-like phase in which neither
confinement nor chiral symmetry breaking is expected.  {}For
values of $N_f$ near but below $11$, the beta function develops a
perturbative infrared stable fixed point at which the trace of the
energy momentum tensor vanishes exactly and the theory is a
non-Abelian conformal field theory.  In this phase, the coupling
constant is small on account of the large number of flavors and so
we do not expect any of the global symmetries to break.  However,
as the number of flavors is decreased, the fixed point becomes
nonperturbative and the coupling constant increases to a critical
value at which chiral symmetry is spontaneously broken.  A
dynamical scale is generated and conformal symmetry is lost.  The
generation of this scale defines the critical number of massless
flavors, i.e. the minimum number of flavors for which the gauge
theory is still conformal and chiral symmetry is still intact.
Below this critical number of flavors, the theory is expected to
confine and the low energy spectrum is hadronic. This discussion
assumes that the conformal and chiral phase transitions coincide,
but whether or not this is true is still controversial. We will
assume here, as corroborated by lattice simulations for $N_c=3$
\cite{mawhinney}, that there is in fact a single conformal/chiral
phase transition. Figure \ref{fig1} summarizes the possible phase
structure for two-color QCD as a function of the number of light
flavors.

\subsection{The Enhanced Global Symmetry Scenario}

When the number of flavors is just below the critical value, the
theory still exhibits chiral symmetry breaking but it is possible
that the vector spectrum changes quite significantly.  In Refs.\
\cite{ARS,DRS}, it was suggested that a new global symmetry may be
dynamically generated.  This symmetry acts on the massive spectrum
of the theory and it is related to the modification of the second
Weinberg spectral function sum rule near the critical number of
flavors \cite{AS}. Indeed, there are examples of supersymmetric
theories with enhanced global symmetry groups \cite{LS}.

{}From eq.\ (\ref{nr2}), one sees that the global symmetry group
becomes $Sp(2N_{f})\times \left[SU(2N_{f})\right] $ for
\begin{equation}
s=4\ ,\qquad h=2 ,
\end{equation}
and the mass splitting between the vectors is zero:
\bea
M_{S}^{2} = g^2(m_V^2 - 2 v^2) = M_X^{2} \ .
\eea
The extra $SU(2N_{f})$ symmetry group acts only on the vector field as
\begin{equation}
A\rightarrow uAu^{\dagger} \ , \qquad u\in [SU(2N_f)] \ ,
\end{equation}
and the effective Lagrangian, eq.\ (\ref{nr2}), reduces to:
\begin{equation}
{\cal L}=v^{2}\,{\rm Tr}\left[ \partial _{\nu }U\partial ^{\nu
}U^{\dagger }\right] +M^{2}{\rm Tr}\left[ A_{\nu }A^{\nu }\right]
\ ,\label{eq:Lenh}
\end{equation}
where $M^{2}=m_V^{2} - 2 v^{2}$ \cite{ARS,DRS}.  Note that the
Lagrangian also possesses an extra global $Z_{2}$ (i.e.
$A\rightarrow zA$, with $z=\pm 1$) symmetry.  If such an enhanced
symmetry emerges, the vectors along the broken generators become
mass degenerate with those along the orthogonal directions. {}For
$N_c=3$ QCD, this corresponds to mass degenerate vector and axial
particles even in the presence of chiral symmetry breaking.

{}From eq.\ (\ref{eq:Lenh}), one finds that in the enhanced
symmetry scenario, the interactions between the vectors and the
Goldstone excitations only appear at the next order in the
derivative counting scheme in the form of the double trace terms
\begin{equation}
{\cal L}_{4}=+a_1 {\rm Tr}\left[\partial_{\rho}U
\partial^{\rho}U^{\dagger}\right]\, {\rm Tr} \left[A_{\nu}
A^{\nu}\right] + a_2 {\rm Tr}\left[\partial_{\rho}U
\partial_{\nu}U^{\dagger}\right]\,{\rm Tr} \left[A^{\rho} A^{\nu}\right] \ ,
\label{foursp}
\end{equation}
where $a_1$ and $a_2$ are real coefficients. If the only extended
symmetry group is the discrete $Z_{2}$ group, then the Lagrangian
can include single trace terms of the form
\begin{equation}
{\rm Tr}\left[\partial_{\rho}U \partial^{\rho}U^{\dagger}A_{\nu}
A^{\nu}\right] \ , \quad {\rm Tr}\left[A_{\rho} A^{\rho} U A_{\nu}^T
A^{T\nu} U^{\dagger} \right] \ .
\end{equation}
The enhanced symmetry scenario imposes very stringent constraints
on the possible form of the $\epsilon$-terms as well.  If we
require that the effective Lagrangian respects the full enhanced
global symmetry, $Sp(2N_{f})\times\left[SU(2N_{f})\right]\times Z_2$,
then there are no vector axial $\epsilon$-terms.  However, if only
the discrete $Z_{2}$ symmetry is imposed, then the surviving
terms are $C_2$, $C_3$, $C_4$, $C_5$, $C_9$ and $C_{10}$.

As supported by ordinary QCD phenomenology \cite{KRS}, we
conjecture the following phase structure before entering
the conformal phase:
\begin{itemize}
\item[$\bullet$]{Approximate local chiral symmetry for small $N_f$}
\item[$\bullet$]{Parity doubling and an extra global symmetry near the critical $N_f$.}
\end{itemize}
{}For a fixed, nonzero chemical potential, the phase structure as
the number of light flavors is increased should be even richer.
{}For instance, when $N_f >11$, the theory is no longer
asymptotically free and the low energy theory is simply the QCD
Lagrangian.  This regime is clearly not in the same universality
class as the one in which the lowest excitations are Goldstone
bosons. {}For $N_f$ smaller than but near the critical value, $N_f
\simeq 8$, we approach the conformal phase. The hadronic mass
scale of the theory vanishes exponentially fast \cite{AS,ATW,FSS}
and all of the physical states whose masses are linked to the
hadronic mass scale, i.e. all non-Goldstone excitations, become
very light.  When there is a large mass gap between the Goldstone
excitations and the rest of the spectrum, the phase diagram for
$N_c=2$ can be predicted using chiral perturbation theory
\cite{KST,KSTVZ,Splittorff:2001mm}.  Near the conformal point,
however, this approach is expected to break down since there is
now a tower of light, non-Goldstone excitations.  This tower of
light states should suppress the formation of both the
quark-antiquark and the diquark condensates by virtue of
conformality. This behavior can be distinguished from the
corrections to the mean--field analysis of the effective
Lagrangian since the latter will not lead to a vanishing of either
condensate.  This has already been verified at next--to--leading
order in Ref.\ \cite{Splittorff:2001fy}. We suggest then that by
measuring diquark condensation at nonzero chemical potential
lattice simulations may be able to single out important features
of the conformal phase transition.


\section{Nonzero Chemical Potential} \label{sec:baryonden}

In this section, we review the procedure for introducing a
chemical potential associated with a conserved charge into the
effective theory.  {}For simplicity, we consider only a baryon
chemical potential, but the generalization to include an isospin
chemical potential is straightforward.  This procedure uniquely
fixes the coefficients of the chemical potential terms in the
effective Lagrangian and is equivalent to the approaches used in
Refs.\ \cite{KST,KSTVZ} to introduce a chemical potential via an
auxiliary extended gauge symmetry.

At nonzero chemical potential, the microscopic Lagrangian has the
form
\begin{equation}
{\cal L}_{N_c=2}=-\frac{1}{4g^2} \vec{G}_{\mu\nu}\cdot\vec{G}^{\mu\nu}
+i\, \bar{Q} \bar{\sigma}^{\nu}\left[\partial_{\nu} - i \mu_B \, B
\, \delta_{0\nu}- i\, \vec{G_{\nu}}\cdot \frac{\vec{\tau}}{2}
\right] Q - \frac{1}{2}m_{q}Q^T \tau_2 \, E \, Q  + {\rm h.c.} \ .
\end{equation}
where the $2 N_f\times 2N_f$ matrix
\bea
B = \frac{1}{2}\left(
\begin{array}{cc}
{\bf 1} & {\bf 0} \\ {\bf 0} & {\bf -1}
\end{array}
\right)
\eea
is the baryon charge matrix for the
quarks and the conjugate quarks\footnote{We adopt the convention of
Ref.\ \cite{Splittorff:2001mm} where the diquarks are chosen to have baryon charge 1.}.
One may check that when written in the basis of the usual $SU(2N_f)$ spinors
this term gives the usual coupling of the quarks to the chemical potential,
$\mu_B {\bar \psi} \gamma_0 \psi$ \cite{KSTVZ}.

After defining $B_{\nu} \equiv \mu_B\, B \,\delta_{0\nu}$, this
Lagrangian is formally invariant under the following $SU(2N_f)$
transformation
\begin{eqnarray}
Q & \rightarrow & u Q \nn \\ B_{\nu} & \rightarrow & u B_{\nu}
u^\dagger - \frac{1}{\mu} u(\d_\nu u^\dagger) \nn \\ E &
\rightarrow &  u^* E u^\dagger \ ,
\end{eqnarray}
and $u\in SU(2N_f)$. Implementing the previous transformations at
the effective Lagrangian, the authors of Refs.\
\cite{KST,KSTVZ,Splittorff:2001mm} were able to uniquely determine the coupling
of the chemical potential to the Goldstone bosons in the effective
theory. The result is that the chemical potential enters into the
effective Lagrangian in the form of a covariant derivative which
we give schematically by
\begin{equation} \label{eq:chemcod}
\partial_\nu \rightarrow \partial_\nu - iB_{\nu}
\ee

This same result can be arrived at by examining the conserved
charges in the effective theory.  The chemical potential is
associated with the conserved charge of baryon number.  Since the
Lagrangian is invariant under a global $U(1)_B$ symmetry group,
there is a conserved current, the zeroth component of which is the
conserved charged associated with the baryon number.  In the
operator formalism, the grand canonical partition function is \bea
Z(\beta,\mu) = {\rm Tr} \, e^{-\beta(\hat{H}-\mu \hat{Q})} \,\, .
\eea Converting from the operator formalism to the path integral
formalism, one finds that after integrating over the field
momentum the introduction of a chemical potential, for a scalar
degree of freedom, serves only to shift the time derivatives in
the fashion: \bea \frac{\partial^2}{\partial t^2} \rightarrow
\frac{\partial^2}{\partial t^2}
        -2 \mu \frac{\partial}{\partial t} + \mu^2 \,\, .
\eea
Note that this is equivalent to replacing the usual derivatives by the
covariant derivative in eq.\ (\ref{eq:chemcod}).

Before discussing the effects of a nonzero chemical potential on
the vectors and the $\epsilon$-terms, it is instructive to review
the predictions of the effective Lagrangian including only the
pseudo-Goldstone excitations \cite{KST,KSTVZ}.  The effects of a
nonzero baryon chemical potential are manifest even at the level
of the chiral effective Lagrangian since $N_f(N_f-1)$ of the
Goldstone modes are diquarks which have nonzero baryon charge.
This is not the case for $N_c=3$ QCD since there is no coupling
between the Goldstone modes and the baryon chemical potential in
chiral perturbation theory. Another related, but salient,
difference is that for $N_c=2$, the critical chemical potential at
which the baryon density becomes nonzero is well within the range
of validity of the effective theory.

Since a nonzero baryon chemical potential only preserves a
$Sp(N_f)\times Sp(N_f)$ subgroup of the original $SU(2N_f)$ symmetry
group, the chemical potential is in competition with the pion mass for
the vacuum structure. It is crucial that this competition be allowed
to take place in the low energy effective Lagrangian.  At zero baryon
chemical potential and nonzero pion mass $Sp(2N_f)$ is left invariant
by the vacuum and the parameterization is given by eqs.\ (\ref{E}) and
(\ref{parametrizationI}). However, if $\mu_B$ exceeds
$m_\pi$, a nonzero diquark condensate is expected to form.  The combination
of the chiral condensate and the diquark condensate leaves only a $Sp(N_f)$
invariance, and consequently one must introduce a parameterization
which allows for this additional symmetry breaking.  A general
parameterization is
\beq U =
e^{i\frac{\Pi^{i}X^{i}}{v}} \overline{\Sigma} \label{U} \ ,
\eeq
where $X_i$ are the broken generators with respect to
$\overline{\Sigma}$. At this point, we consider only an even number of
flavors and following Ref.\ \cite{KSTVZ} we introduce
\beq \overline{\Sigma} = E
\cos(\al) + D \sin(\al) \ , \label{ansatz} \eeq
where
\beqn
\label{eq:Dmatrix} D \equiv i\,\left(\begin{array}{cc} {\cal I}
&{\bf 0} \\ {\bf 0} & {\cal I}
\end{array}\right)
 \,\,\,\,\,\, {\rm with} \,\,\,\,\,\, {\cal I} \equiv  \left(\begin{array}{cc} 0 &
     -{\bf 1} \\ {\bf 1} & 0 \end{array}\right)  \ .
\eeqn
The variational parameter $\al$ is determined by minimizing the free energy.

After introducing the chemical potential as discussed above, the most general
effective Lagrangian containing only pseudo-Goldstone excitations to
second order in the chiral counting is
\beqn {\cal L}(U) & = & v^2\Tr[\d_\nu U\d^\nu
U^\dagger]-i4\mu_B v^2\Tr[BU \d_0U^\dagger]
+ 2 v^2\mu_B^2\left(\Tr[BU B^{\rm T}U^\dagger]+\Tr[BB]\right) \nn \\
 &+&v^2m_\pi^2\Tr[{\cal M}U+{\cal M}^\dagger U^\dagger] \ .
\label{Lmu1} \eeqn The phase diagram and dispersion relations for
the pseudo-Goldstone modes were derived in \cite{KST,KSTVZ}. The
steps leading to the phase diagram are as follows. {}First, one
extremizes the stationary action \beqn {\cal L}(\overline{\Sigma})
& = & 2 v^2 \mu_{B}^2 \left( \Tr[B\overline{\Sigma}
B^{\rm T}\overline{\Sigma}^\dagger]+\Tr[BB]\right)
+v^2m_\pi^2\Tr[{\cal M}\overline{\Sigma}+{\cal M}^\dagger
\overline{\Sigma}^\dagger]
 \label{stL}
\eeqn with respect to $\al$. This leads to the following
nonanalytic behavior in $\al$: \beqn \al=0 & \ \ \ \  {\rm for} &
\ \ \  \mu_{B} < m_\pi \\ \cos(\al) = \frac{m_\pi^2}{\mu_{B}^2} &
\ \ \ \  {\rm for} &  \ \ \ \mu_{B}\geq m_\pi \ . \label{muc}
\eeqn It was shown in Ref.\ \cite{KSTVZ} that this vacuum
direction does indeed parameterize the global minimum of the
static potential ${\cal L}(\overline{\Sigma})$.  The condensates
and densities are simply given by derivatives of the static
potential with respect to the appropriate sources.  {}For the
reader's convenience, we repeat these expressions along with the
baryon density \beqn \langle\bar{\psi}\psi\rangle = 2 N_f G
\cos\al \ , \ \ \ \langle\psi\psi\rangle = 2 N_f G \sin\al \ , \ \
\  n_B = 8 N_fv^2\mu_B\sin^2\al \ . \label{condensates} \eeqn The
constant $G$ is the chiral condensate in the chiral limit.

We want to stress that the above analysis assumes that the vector
spectrum is heavy as compared to the mass of the pseudo-Goldstone
excitations. This assumption, however should not hold for large
enough number of flavors. Hence it would be very interesting to
see if lattice calculations find deviations for $N_f\simeq 8$. We
also remark that the Lagrangian in eq.\ (\ref{Lmu1}) is investigated at
tree-level.


\section{The $\epsilon$ Terms at Nonzero Chemical Potential}
\label{sec:e-terms-mu}

In this section we compute the effect of the chemical potential in
the $\epsilon$-terms. {}For simplicity we neglect the spin-one
fields and consider only $N_f=2$.

\subsection{Nonzero baryon chemical potential}

The effect of the baryon chemical potential through the
$\epsilon$-terms is obtained by substituting $A_\nu$ by
$B_\nu=\mu_B \delta_{\nu,0} B$  in eq.\ (\ref{generalWZ}) with
$C_i=C$, for $i=1,\ldots,10$ and $C_{11}=0$. Since  $B_\nu$ only
has a temporal component, all terms with more than one $B_\nu$
vanish on account of the antisymmetry of the differential forms.
The resulting term is
\begin{equation}
\Gamma[U,B] ~ = ~ i\, \mu_B \, 10 \,C\,\int_{M^{4}}{\rm Tr}\left[ B\alpha
^{3}\right]=i\, \mu_B \, 10\,C\,
\int_{M^4}\Tr\left[B\alpha_{i}\alpha_{j}\alpha_{k}\right]\epsilon^{0ijk}
d^4\,x \ . \label{UniqueLT}
\end{equation}
In general, this term does not vanish and it is instructive to
investigate it in more detail. By expanding $\alpha$ to first
order in derivatives and lowest order in Goldstone fields, we have
the Lagrangian density
\begin{equation}
i\,\mu_B \,10\,C\,
\Tr\left[B\alpha_{i}\alpha_{j}\alpha_{k}\right]\epsilon^{0ijk}=
5\,\frac{C \mu_B}{v^3} \, \epsilon^{0ijk} {\rm Tr}
\left[B\,\left[X^l,X^m\right]\,X^n\right] \partial_i \pi^l
\partial_j \pi^m \partial_k \pi^n + \cdots \ . \label{bho}
\end{equation}
{}For $N_f=2$ the trace is nonzero only if $l,m,n=1,2,3$, that is
only the pion generators, $X^i$ with $i=1,2,3$ contribute to the
trace (see the Appendix for the conventions). An explicit
calculation yields
\begin{equation}
i\,\mu_B\,10\,C\,\int_{M^{4}}{\rm Tr}\left[ B\alpha
^{3}\right]=i\,5\,\frac{C\mu_B}{4\sqrt{2}\,v^3}
\int_{M^{4}}\epsilon^{lmn}\epsilon^{0ijk}
\partial_i \pi^l \partial_j \pi^m \partial_k \pi^n d^4x +\cdots \ .
\end{equation}
The integral is related to the winding number when considering
nontrivial topological sectors of the theory and it is naturally
coupled to $\mu_B$.

In the evaluation of eq.\ (\ref{UniqueLT}), we used the explicit
representation of the generators given in eq.\ (\ref{generators}).
As the diquark condensation sets in and the minimum of the
stationary Lagrangian rotates according to eq.\ (\ref{ansatz}),
this representation breaks down. If, however, we choose to write
the rotation of the generators explicitly then we may use the
original representation of the generators at the cost of rotating
the sources \cite{KSTVZ}: \beq B\to B \cos\varphi - BDE
\sin\varphi \ . \eeq {}For $N_f=2$ one can verify that \beq
\epsilon^{0ijk} {\rm Tr} \left[BDE\, X^l\,X^m\,X^n\right]
\partial_i \pi^l
\partial_j \pi^m \partial_k \pi^n ~ = ~ 0 \ ,
\eeq 
and the final result is 
\beqn i\,\mu_B\, 10\,C\,\int_{M^{4}}{\rm
Tr}\left[ B\alpha ^{3}\right] & = & i\,5\,\cos(\varphi)\frac{C \,
\mu_B}{4\sqrt{2}\,v^3}
\sum_{l,m,n=1}^{3}\int_{M^{4}}\epsilon^{lmn}\epsilon^{0ijk}
\partial_i \pi^l \partial_j \pi^m \partial_k \pi^n d^4x +\cdots\ ,
\eeqn where the sum over $l,m,n$ is performed for $l,m,n=1,2,3$.
Note that the contribution from the winding number to the baryon
density is proportional to $\cos(\varphi)$ and hence decreases with
increasing $\mu_B$ like $m_\pi^2/\mu_B$.

\subsection{Nonzero isospin chemical potential}

Above we observed that the $\epsilon$ terms couple to the baryon
chemical potential via the pions transforming according to the adjoint
representation of $SU_V(N_f)$.  The diquark sector, however, was not
active. The situation is reversed when considering a nonzero
isospin chemical potential. Since the quarks carry different flavor
quantum numbers,
it is possible and even natural to introduce different chemical
potentials for the different flavors. Let us consider $N_f=2$ and
introduce the chemical potentials
\begin{equation}
\mu_B\equiv\mu_u+\mu_d \quad  {\rm and} \quad
\mu_I\equiv\mu_u-\mu_d \ .
\end{equation}
Here we consider $\mu_B=0$.  The effects of $\mu_I$ in the
$\epsilon$ terms enters through the substitution in eq.\
(\ref{generalWZ}) of $A_\nu$ by \beq I_\nu \equiv \mu_I
\delta_{0\nu} I \equiv \mu_I \delta_{0\nu}
\frac{1}{2}\left(\begin{array}{cc} {\tau^3} &{\bf 0} \\ {\bf 0} &
{-\tau^3} \end{array}\right)  \ . \eeq The microscopic two-color
Lagrangian is invariant under the combined exchanges
$\mu_I\leftrightarrow \mu_B$ and $d_L\leftrightarrow
i\sigma_2\tau_2 d_R^*$ where $\sigma_2$ and $\tau_2$ are Pauli
matrices acting in spin and color space respectively. ($d_L$ is
the left down-quark field and the combination $i\sigma_2\tau_2
d_R^*$ is known as the conjugate quark state.) This invariance is
inherited by the effective Lagrangian where it translates into
$\mu_I\leftrightarrow \mu_B$ and
$(\pi_1X_1,\pi_2X_2)\leftrightarrow (\pi_4X_4,\pi_5X_5)$. Using
this we conclude from the explicit calculation in the case of
baryon chemical potential that for $N_f=$2 and $\mu_B=0$ \beqn
i\, \mu_I \, 10\,C\, \int_{M^{4}}{\rm Tr}\left[ I\alpha ^{3}\right] & = &
i\, 5\,\cos(\varphi)\frac{C\,\mu_I}{4\sqrt{2}\,v^3}
\sum_{l,m,n=3}^{5}\int_{M^{4}}\epsilon^{lmn}\epsilon^{0ijk}
\partial_i \pi^l \partial_j \pi^m \partial_k \pi^n d^4x+\ldots \ ,
\eeqn
where now the sum over the flavor indices is performed for
$l,m,n=3,4,5$, while $\cos\varphi=(m_\pi/\mu_I)^2$ for
$\mu_I>m_{\pi}$ and $\varphi=0$ otherwise.
As above, this term is also related to the winding number of the
diquark Goldstone fields.
We note that in these $\epsilon$-terms $\mu_B$ couples only to the
isospin triplet $\pi^1,\pi^2,\pi^3$, while $\mu_I$ couples only to the
baryon triplet $\pi^3,\pi^4,\pi^5$.  This should be contrasted with
three-color QCD in which the baryon chemical potential couples to the
isospin triplet, but there is no coupling at all between the isospin
chemical potential and the baryonic sector of the theory.  For this
reason, the spectrum of solitons in two-color QCD is richer than for
$N_c=3$.  To our knowledge this exciting possibility has not been
explored.  Note that it is straightforward to extend these results
into the $(\mu_B,\mu_I)$-plane, using the results of \cite{Splittorff:2001mm}. As a
direct application of eq.\ (\ref{UniqueLT}), one can also extend these
results to higher $N_f$.

Finally we recall that a relevant feature of the Wess-Zumino term
is that is required to saturate at the effective Lagrangian level
the 't Hooft global anomaly conditions. It is also well known that
the Goldstone bosons are sufficient (when chiral symmetry is
spontaneously broken) to saturate the anomaly matching conditions.
So no other light degree of freedom close to the conformal phase
transition is expected to affect the Wess-Zumino term. In the
conformal region the underlying quarks will automatically saturate
the 't Hooft anomaly conditions. So it would certainly be
interesting to monitor these terms close to the conformal phase
transition.

\section{The Vector Spectrum at Nonzero Chemical Potential}
\label{sec:vec-and-mu}

We now examine the effects of a nonzero baryon chemical potential
on the spectrum of vectors.  The chemical potential enters simply
by modifying the derivatives acting on the vector fields:
\begin{equation}
\partial_{\nu} A_{\rho} \rightarrow \partial_{\nu}A_{\rho} - i
\left[B_{\nu}\ ,A_{\rho}\right]\ ,
\end{equation}
with $B_{\nu}=\mu_B \,\delta_{\nu 0} B\equiv V_{\nu} B$ where
$V=(\mu_B \ ,\vec{0})$. Using the previous prescription for the
vector kinetic term, we find
\begin{eqnarray}
{\rm Tr} \left[F_{\rho \nu} F^{\rho \nu}\right] \rightarrow {\rm
Tr} \left[F_{\rho \nu} F^{\rho \nu}\right] - 4i{\rm Tr}
\left[F_{\rho\nu}\left[B^{\rho},A^{\nu}\right]\right] - 2 {\rm Tr}
\left[\left[B_{\rho},A_{\nu}\right]\left[B^{\rho},A^{\nu}\right] -
\left[B_{\rho},A_{\nu}\right]\left[B^{\nu},A^{\rho}\right]\right]
\ .
\end{eqnarray}
Retaining only the quadratic terms in eq.\ (\ref{eq:kin}) and
integrating by parts yields
\begin{eqnarray}
{\cal L}_{quad}&=&\frac{1}{2g^2}A_{\rho}^a\Big\{ \delta_{ab}
\Big[\eta^{\rho\nu}\Box-\partial^{\rho}\partial^{\nu}\Big] -
4i\gamma_{ab}\left[\eta^{\rho \nu}V\cdot \partial -
\frac{1}{2}(V^{\rho}
\partial^{\nu} + V^{\nu}
\partial^{\rho})\right] \\ \nonumber 
&+& 2
\chi_{ab}\left[V\cdot V \eta^{\rho\nu}-V^{\rho}V^{\nu}\right]\Big\}
A_{\nu}^b
\end{eqnarray}
with
\begin{equation}
\gamma_{ab}={\rm Tr}\left[T^a\left[B,T^b\right] \right]\ , \qquad
\chi_{ab}={\rm
Tr}\left[\left[B,T^a\right]\left[B,T^b\right]\right] \ .
\end{equation}
Note that the inclusion of a baryon chemical potential
induces a ``magnetic-type'' mass term for the vectors
at tree-level.  To complete the
quadratic terms, we include the ordinary mass term already
present in eq.~(\ref{nr2}):
\begin{equation}
{\cal L}_{mass}=\frac{v^2}{2}A^a_{\rho}\eta^{\rho\nu}\xi_{ab}A_{\nu}^b \
,
\end{equation}
with
\begin{equation}
\xi_{ab}=\frac{1}{2v^2g^2}\left\{\left(M_{S}^2+M_{X}^2\right)\delta_{ab}+2\left(M_X^2-M_S^2\right){\rm
Tr}\left[T^a\overline{\Sigma}{T^b}^T\overline{\Sigma}^{\dagger}\right]\right\}
\ . \label{mixed}
\end{equation}
This term gives the lowest order interaction between the Goldstone
sector and the vectors. We assume that the vacuum alignment is fixed
by the Goldstone sector and so this interaction term gives the leading
order effect of the alignment on the dispersion relations for the
vectors\footnote{We remind the reader that at the tree level there is
also a quadratic term of the type $\partial^{\mu}\Pi^a A^a_{\mu}$.
This mixing term can be diagonalized \cite{ARS,KS} away by the
field redefinition $A\rightarrow A+v\frac{4-s}{2M^2_A}
\partial\Pi$ while leaving the mass spectrum unchanged.  $A$
stands for the vectors which mix with the pions while $M^2_A$ is
its tree level mass.}.

Having extracted the quadratic terms for the vectors, we can
calculate the mass gap, i.e. the zero momentum limit of
the dispersion relations.  In this limit, the temporal
components have no energy dependence while the
quadratic part of the spatial components has the form
\begin{eqnarray}
A_i^a\left[\delta_{ab}E^2-4\gamma_{ab}\mu_B E -g^2v^2\xi_{ab} -2
\mu_B^2\chi_{ab}\right]A_i^b \ .
\label{genVmass}
\end{eqnarray}
Up until this point $N_f$ is completely general and the masses of the
$(2N_f)^2-1$ vector modes can be obtained by solving for $E$
\be
\det \left[\delta_{ab}E^2-4\gamma_{ab}\mu_B E -g^2v^2\xi_{ab} -2
\mu_B^2\chi_{ab}\right] = 0 \ .
\label{detGen}
\ee
Here we choose to focous on $N_f=2$. The basis of the 15 vectors is taken as
that given in the Appendix. We choose the following ordering,
$A_i^a=A_i^aS^a$ for $a=1,\ldots,10$, $A_i^a=A_i^aX^{a-10}$ for $a=11,\ldots,15$.
Explicitly calculating the traces in (\ref{detGen}) using the said basis one
finds that the zero
momentum propagator matrix is block diagonal with three $1\times 1$-terms and
four $3\times 3$ blocks. The
diagonal terms are for $A_i^1$, $A_i^2$, and $A_i^3$. The masses in
these channels take the single value
\be
M^a=M_S  \ \ , \ \ \ a=1,2,3 \ .
\ee
The  $3\times 3$ blocks
mix ($A_i^5$,$A_i^6$,$A_i^{12}$), ($A_i^7$,$A_i^8$,$A_i^{11}$),
($A_i^9$,$A_i^{10}$,$A_i^{13}$), and  ($A_i^{14}$,$A_i^{15}$,$A_i^{4}$)
respectively. The first three triplets are degenerate and the masses are
obtained by solving for $E$ 
\be
\left|\begin{array}{ccc} E^2 + \mu^2 -M_S^2 & 2 i E \mu & 0 \\
 & & \\
- 2 i E \mu & E^2 + \mu^2 -\frac{M_X^2+M_S^2}{2}
+\frac{M_X^2-M_S^2}{2}\cos(2\varphi) & \frac{M_X^2-M_S^2}{2}\sin(2\varphi) \\ 

& & \\
0 &\frac{M_X^2-M_S^2}{2}\sin(2\varphi) & E^2 - \frac{M_X^2+M_S^2}{2} -
\frac{M_X^2-M_S^2}{2}\cos(2\varphi) \end{array}\right| =0 
\ ,
\ee
while in the ($A_i^{14}$,$A_i^{15}$,$A_i^{4}$) sector we must solve
\be
\left|\begin{array}{ccc}
    E^2-\frac{M_X^2+M_S^2}{2}+\frac{M_X^2-M_S^2}{2}\cos(2\varphi)  & 0 &
    \frac{M_X^2-M_S^2}{2}\sin(2\varphi) \\ & & \\ 
0 & E^2+\mu^2-\frac{M_X^2+M_S^2}{2}-\frac{M_X^2-M_S^2}{2}\cos(2\varphi) & 2 i E \mu \\
& & \\ 
 \frac{M_X^2-M_S^2}{2}\sin(2\varphi) & -2 i E \mu & E^2+\mu^2-M_X^2 
\end{array}\right| =0 
\ .
\ee

\begin{figure}[htb]
\begin{center}
\leavevmode {\epsfxsize=12.0truecm \epsfbox{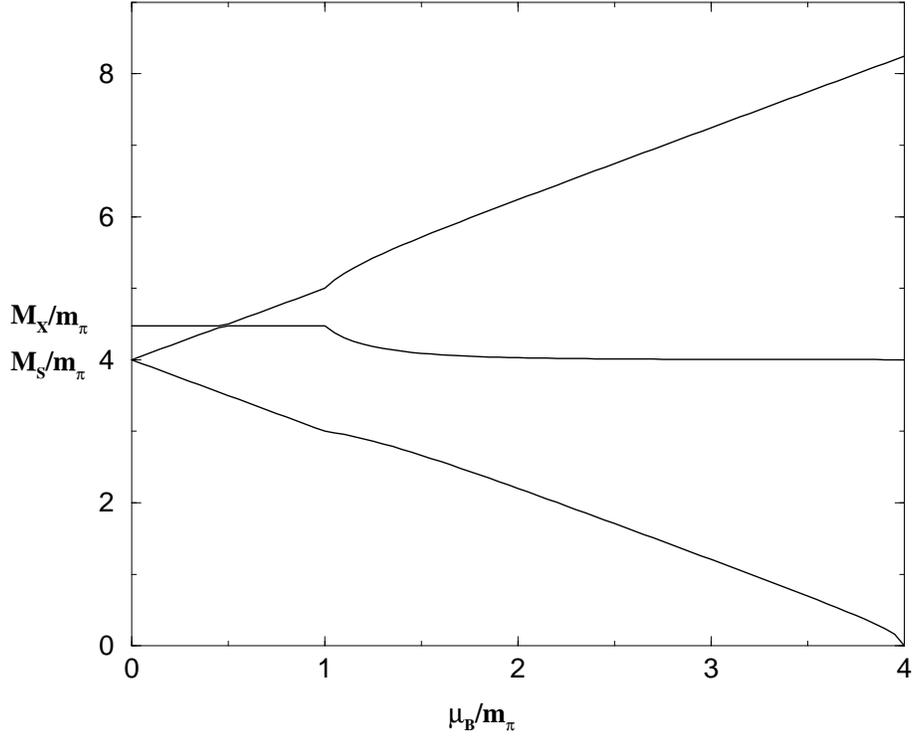}}
\end{center}
\caption{The triply degenerate masses of the vectors in the sectors
($A_i^5$,$A_i^6$,$A_i^{12}$), ($A_i^7$,$A_i^8$,$A_i^{11}$),  and
($A_i^9$,$A_i^{10}$,$A_i^{13}$). We have choosen $h=s=0$, $m_V=4m_\pi$
$v=m_\pi$ and $g=1$ to make the plot. The choice $h=s=0$ realizes the limit
where the vector Lagrangian only breaks local chiral symmetry through the term
proportional to $m_V^2$.}\label{fig:3triplets}
\end{figure}

\begin{figure}[htb]
\begin{center}
\leavevmode {\epsfxsize=12.0truecm \epsfbox{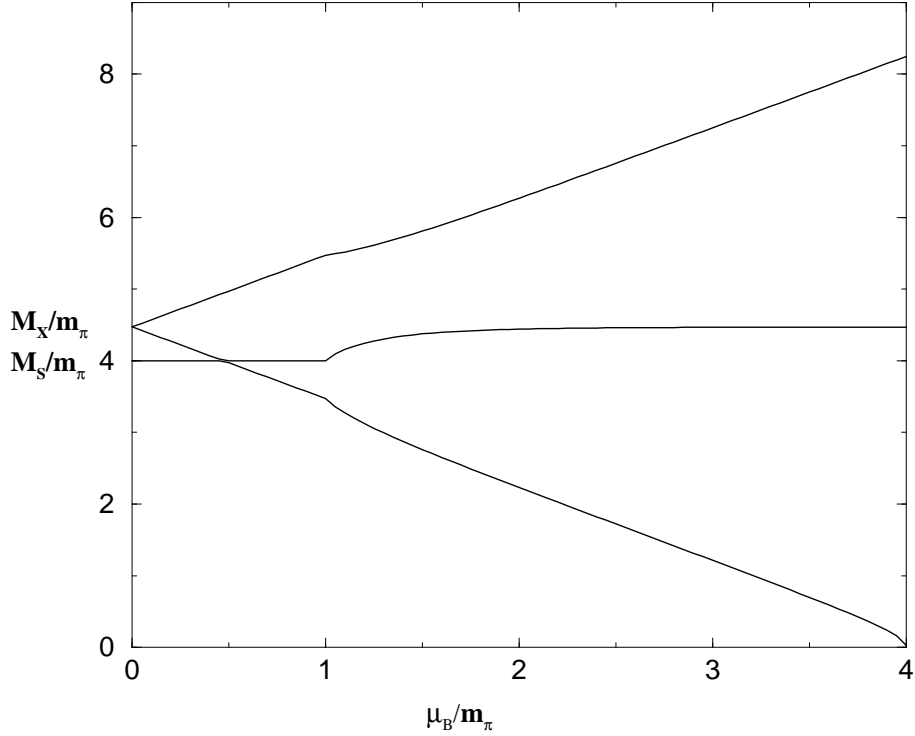}}
\end{center}
\caption{The masses of the three vectors in the sector
($A_i^{14}$,$A_i^{15}$,$A_i^{4}$) with the same choice of the parameters
$h,s,m_V,v,g$ as in figure \ref{fig:3triplets}. Note the square-root singularity of
the lightest vector mass just below the critical chemical potential in
accordence with eq.(\ref{MII}). }\label{fig:1triplet}
\end{figure}

{}To illustrate the calculation of the general zero momentum propagator
matrix, we now investigate the ``$X-$type'' diquark vector states,
i.e. the fields $A_i^{14}$ and $A_i^{15}$. The $\gamma_{ab}$ and the
$\xi_{ab}$ terms do not mix $A_i^{14}$ and $A_i^{15}$ with the
rest of the 15 $A_i^a$ fields. However, the $A_\nu^{14}$ and
$A_\nu^{15}$ states mix since $[B,X^{4}]=-iX^{5}$ and $[B,X^{5}]=+
iX^{4}$ and so
\begin{equation}
\gamma_{14,15}=-\gamma_{15,14}=\frac{i}{2} \ , \qquad
\chi_{ab}=-\frac{\delta_{ab}}{2}
\end{equation}
with $a,b=14,15$. Additionally, the last term in eq.~(\ref{mixed}) depends on
$\overline{\Sigma}$ and it mixes $A_i^{14}$ and $A_i^{15}$ with $A_i^4$.
This mixing, however, vanishes in two cases:
\begin{itemize}
\item
[({\bf i})]{the nonsuperfluid phase, i.e. where $\mu_B < m_{\pi}$
with $\varphi=0$ and $\overline{\Sigma}=E$,}
\item[({\bf ii})]{the superfluid phase with  $\mu_B \gg m_{\pi}$ where
$\varphi\approx \pi/2$ and $\overline{\Sigma}\approx D$.}
\end{itemize}
In case ({\bf i}), we find
\begin{eqnarray}
\xi_{ab}&=& \frac{M_X^2}{v^2}\delta_{ab} \qquad
\varphi=0 \ , \eea with $a,b=4,5$, while for case ({\bf ii}) we
have \bea
 \xi_{14,14}&=&\frac{M_S^2}{v^2}~~~ \qquad
\varphi=\frac{\pi}{2} \ , \\ \xi_{15,15}&=&
\frac{M_X^2}{v^2}~~~ \qquad
\varphi=\frac{\pi}{2}\  .
\end{eqnarray}
After diagonalizing the quadratic mass terms, the mass gap in
case ({\bf i}) is
\begin{eqnarray}
M_{V^{\pm}}&=& M_X \pm \mu_B \ .
\end{eqnarray}
Here $V^{\pm}$ labels the states which diagonalize the mass
matrix.
In case ({\bf ii}), we find that
\be
\label{MII}
M_{V^{\pm}}^2 =  \frac{1}{2}\Bigg[M_S^2+M_X^2+ 2\mu_B^2 \pm
\sqrt{ (M_S^2-M_X^2)^2+8\mu_B^2(M_S^2+M_X^2)} \Bigg] \ ,
\ee
at  $\varphi=\pi/2$.
{}For $\mu_B=M_S$ or $M_X$, $V^-$ becomes massless. This
suggests that vectors condense for $\mu_B={\rm min}\{M_S,M_X\}$
breaking rotational invariance. At this value of $\mu_B$, the
approach breaks down since the effects of such a condensation is
not accounted for in eq.\ (\ref{ansatz}). This calculation is in agreement
with the general solutions plotted in figure \ref{fig:3triplets} and in
\ref{fig:1triplet}. In these plots we have chosen $h=s=0$, $m_V=4m_\pi$
$v=m_\pi$ and $g=1$. This choice leads to the conventional vector mass
splitting between the ``X''-like and the ``S''-like vectors at zero chemical
potential.  As the triplets consist of two ``S''-like and one ``X''-like
generator or of two ``X''-like and one ``S''-like generator this mass
splitting is apparent on the plots. The ``X''-like and ``S''-like vectors
only mix for $0<\varphi<\pi/2$, that is only for $\mu_B>m_\pi$ and
$\mu_B\not\gg m_\pi$.

The mass gaps of the vectors and the possibility of vector
condensation can be studied in lattice simulations.  Such studies
would give direct information about the conformal phase transition
since the masses of the vectors depend directly on the parameters
$s$, $h$, and $m_V$.  {}Fixing these parameters is sufficient to
check our conjecture that for a small number of light flavors we
can use an approximate local chiral gauge theory to introduce the
vector-Goldstone interactions (i.e. $h=s=0$ but $m_V^2\neq0$), while
for a large number of flavors an enhanced global symmetry (i.e. $s=2h=4$) 
may emerge. We hope that lattice simulations can shed light on this
issue in the near future.


\section{The Two-Color Linear Effective Lagrangian}
\label{sec:linrev}

In this section, we study the effective Lagrangian for two-color QCD
for which the chiral symmetry is linearly realized.  The authors of
Ref.\ \cite{Rapp:1998zu} used this theory to study the superfluid
phase transition at nonzero baryon chemical potential and zero isospin
chemical potential.  This theory was also used in Ref.\ \cite{Wirstam}
to examine the universal properties of the chiral symmetry restoring
phase transition of two-color QCD at nonzero temperature and vanishing
chemical potential.  The random matrix model for two-color QCD at
nonzero chemical potential considered by the authors of
Ref. \cite{Vanderheyden:2001gx} also bears resemblence to the linear
sigma model considered here.  We begin by introducing the effective
Lagrangian including the couplings between the pseudo-Goldstone
excitations and the spin-one sector. We then derive the phase diagram
at nonzero baryon and isospin chemical potential.

In the linear effective Lagrangian, the $2N_f^2-N_f-1$ Goldstone fields
which are present in the nonlinear effective Lagrangian are accompanied by
a scalar particle, $\sigma$. These fields are elements of the
antisymmetric $2N_f\times2N_f$ matrix
\beq
M = \frac{1}{2\sqrt{2}} \left( \sigma -i 2 \sqrt{2} \pi^{a} \,
        X^{a}\right)\, E  \label{eq:Mfieldmatrix} \ .
\eeq Under the action of $u\in$\,SU(2$N_f$), $M$ transforms as
\beq M\to uMu^T \ . \eeq A vector field, $A_\nu$, can be
introduced in a fashion similar to that for the nonlinear
effective Lagrangian and so transforms according to eq.\
(\ref{transf-A}). Hence, it is useful to define the covariant
derivative acting on $M$ \beq D_{\nu }M=\partial _{\nu }M-iA_{\nu
}M-iMA_{\nu }^{T} \,\, . \eeq As in the nonlinear realization, we
write a general Lagrangian consistent with the global chiral
symmetry invariance: \beqn \label{linsig11} {\cal L}_{linear} &=&
{\rm Tr}\left[ \partial_{\nu} M
        \partial^{\nu} M^{\dagger}
        \right] - m^2\,{\rm Tr}\left[ M \,M^{\dagger} \right]+
    i c_1 {\rm Tr}\left[A_\nu \left(M \partial^\nu M^{\dagger}-\partial^{\nu}
    M M^{\dagger}\right)\right]
        \\ \nonumber
        &+&
         c_2 {\rm Tr}\left[ A^\nu M A_\nu^{T} M^{\dagger}  \right]+
         c_3 {\rm Tr}\left[A_\nu \, \, A^\nu\right] + c_4 {\rm Tr}\left[A_\nu A^{\nu}
    M M^{\dagger}\right]\\ \nn
        &-& \lambda_1\, {\rm Tr}\left[ M \,M^{\dagger} \right]^2
        - \lambda_2 \, \left({\rm Tr}\left[ M \,M^{\dagger}\right]\right)^2
    -H {\rm Re}\left(
        {\rm Tr}\left[ {\cal M} M\right]\right)/\sqrt{2} \\ \nonumber
    &-&\frac{1}{2g^2} {\rm Tr}\left[F_{\mu\nu}F^{\mu\nu}\right] \,\, ,
\eeqn where ${\cal M}$ is given by eq.\ (\ref{eq:Mfieldmatrix}),
$F_{\mu\nu}$ is given in eq.\ (\ref{eq:F}) and all the
coefficients are real.  Since it is determined by the same
symmetry principle, the linear effective Lagrangian is very
similar in form to the nonlinear version. We have only included
potential terms up to fourth order in the mass dimension. The
Gell--Mann--Oakes--Renner relation fixes the coefficient of the
explicit symmetry breaking term to be $H=m_\pi^2 f_\pi$, where
$m_\pi$ is the degenerate mass of the pseudo--Goldstone particles
at zero baryon and isospin chemical potential and
$f_\pi=2\sqrt{2}\,v$.  {}For simplicity, we are ignoring terms
which account for the explicit breaking of the anomalous $U(1)_A$
symmetry group.

{}For the purposes of deriving the phase diagram, we consider only
$N_f=2$ in the absence of spin-one particles. As discussed above,
the chemical potentials can be introduced by formally replacing
the usual derivatives with the covariant derivative \beq
\partial_\nu M \rightarrow \partial_\nu M - i
(\mu_B B+ \mu_I I)\delta_{0\nu}M - i M (\mu_B B+ \mu_I I)\delta_{0\nu} \ ,
\ee
where
\be
\mu_B B+ \mu_I I=\frac{\mu_B}{2}
\left(\begin{array}{cc} {\bf 1} & \bf 0 \\ \bf 0 & -{\bf 1}
  \end{array}\right)
+\frac{\mu_I}{2}
\left(\begin{array}{cc} {\bf \tau}^3 & \bf 0 \\ \bf 0 & -{\bf \tau}^3
  \end{array}\right)   \ .
\eeq
The linear effective Lagrangian at nonzero baryon and isospin chemical
potential is then
\bea \label{linearsig}
{\cal L}_{linear} &=& {\rm Tr}\left[ \partial_{\nu} M
        \partial^{\nu} M^{\dagger}
        \right] - m^2\, {\rm Tr}\left[ M \,M^{\dagger} \right]
        - 4\,\lambda_1 {\rm Tr}\left[ M \,M^{\dagger} \right]^2
        - 4\,\lambda_2 \left({\rm Tr}\left[ M \,M^{\dagger}\right]\right)^2
        \\ \nonumber
        &+&2 \, {\rm Tr}\left[ M \, (\mu_B\, B+ \mu_I \, I) \, M^{\dagger}
        (\mu_B\, B+ \mu_I \, I) + M\,M^{\dagger}(\mu_B\, B+ \mu_I \, I)^2 \right] \\ \nn
        &+& 4i {\rm Tr}\left[(\mu_B\, B+ \mu_I \, I) \partial_0 M \,M^{\dagger}
        \right]
         -\frac{H}{\sqrt{2}} {\rm Re}\left(
        {\rm Tr}\left[ {\cal M} M\right]\right) \,\, .
\eea {}For $N_f=2$, there is a triplet of pions, a scalar
quark--antiquark bound state, a scalar diquark, and its
antiparticle.  The diquarks carry baryon charge $\pm 1$ and the
triplet of pions have isospin charges $\pm 1$ and $0$. Computing
the traces in ${\cal L}$ for $N_f=2$, one finds the effective
potential \be \label{sigpot} V = \frac{m^2}{2} \left(\sigma^2 +
\pi^a\pi^a\right)
        +\frac{\lambda}{4} \, \left(\sigma^2 + \pi^a\pi^a\right)^2
        -  \frac{\mu_I^2}{2} \,  \left[ (\pi^1)^2+(\pi^2)^2\right] -
        \frac{\mu_B^2}{2} \,  \left[(\pi^4)^2+(\pi^5)^2\right] - H \, \sigma \,\, .
\ee
There is no distinction between the two quartic terms in
eq.\ (\ref{linearsig}) for $N_f=2$, and so the two couplings, $\lambda_1$ and
$\lambda_2$, have been absorbed into a single coupling, $\lambda$.
The coefficients $m$ and $\lambda$ may be expressed in terms of
$m_\pi$, $m_\sigma$ and $f_\pi$:
\begin{mathletters}
\bea
m^2 &=& \frac{3 m_\pi^2-m_\sigma^2}{2} \\
\lambda &=& \frac{m_\sigma^2-m_\pi^2}{2f_\pi^2} \,\, .
\eea
\end{mathletters}
The potential is manifestly invariant under a $O(2) \times O(2)$
symmetry group, i.e. under independent rotations of $(\pi^1,\pi^2)$
and $(\pi^4,\pi^5)$.  An explicit calculation shows that at the level
of the Lagrangian this invariance is not violated by the term linear
in the time derivative:
\bea
4 i {\rm Tr}\left[(\mu_B\, B+ \mu_I \, I)
\partial_0 M\, M^{\dagger}\right] &=& \mu_I (\pi^1 \d_0 \pi^2-\pi^2 \d_0 \pi^1)
        +\mu_B (\pi^5 \d_0 \pi^4-\pi^4 \d_0 \pi^5)  \,\, .
\eea

We denote the pion condensate by $\langle \pi \rangle$,
the diquark condensate by $\langle D \rangle$ and the chiral
condensate by $\langle \sigma \rangle$.  Since $\pi^3$ does not
couple to either chemical potential, it is not expected to condense.  The
$O(2) \times O(2)$ symmetry invariance of the effective potential
requires that
\be
\langle \pi^1 \rangle^2 + \langle \pi^2 \rangle^2  = \langle \pi \rangle^2 \ ,\qquad
\langle \pi^4 \rangle^2 + \langle \pi^5 \rangle^2  = \langle D \rangle^2  \  ,
\ee
in the unbroken phase.
The values of the condensates as a function of the baryon
and isospin chemical potentials are found by extremizing
the effective potential:
\begin{mathletters}
\bea
\frac{\delta V}{\delta\langle \sigma \rangle} &=& 0 = -H +
        \langle \sigma \rangle \left[ m^2+\lambda
        \left(\langle \sigma \rangle^2 +\pimvev^2  + \langle D \rangle^2\right)
        \right] \\
\frac{\delta V}{\delta\pimvev} &=& 0 =
          \pimvev  \left[ m^2-\mu_I^2+\lambda
        \left(\langle \sigma \rangle^2 + \pimvev^2+  \langle D \rangle^2\right)
        \right] \\
\frac{\delta V}{\delta\langle D \rangle} &=& 0 =
        \Svev \left[ m^2-\mu_B^2+\lambda
        \left(\langle \sigma \rangle^2+  \pimvev^2+  \langle D \rangle^2\right)
        \right] \,\, .
\eea
\end{mathletters}
There are three real solutions to this set of equations
corresponding to three distinct phases.  The first solution
is given by $\pimvev=\Svev=0$ with $\sigvev$ determined by
the nonzero solution to $H = m_\pi^2 f_\pi = \sigvev \left(m^2 + \lambda \sigvev\right)$.
In this phase, $\langle \sigma \rangle$ is constant, i.e.
equal to $f_\pi$.
The other two solutions are
\begin{mathletters}
\bea
\langle \sigma \rangle &=& \frac{H}{\mu_B^2} \\
\langle D \rangle &=& \sqrt{\frac{\mu_B^2-m^2-\frac{\lambda H^2}
        {\mu_B^4}}{\lambda}} \\
\pimvev &=& 0\,\, ,
\eea
\end{mathletters}
and
\begin{mathletters}
\bea
\langle \sigma \rangle &=& \frac{H}{\mu_I^2} \\
\langle D \rangle &=& 0\\
\pimvev &=& \sqrt{\frac{\mu_I^2-m^2-\frac{\lambda H^2}
        {\mu_I^4}}{\lambda}} \,\, .
\eea
\end{mathletters}
Note that even for nonzero quark masses,
$\pimvev$ and $\Svev$ are true order parameters.  The phase boundaries in
the $(\mu_I,\mu_B)$ plane can now be calculated and are given by
solving the inequalities
\begin{mathletters}
\be
\frac{\delta^2 V}{\delta \langle D \rangle \delta \langle D \rangle}(\sigvev,\pimvev,0)
        > 0  \,\, ,
\ee
and
\be
\frac{\delta^2 V}{\delta \pimvev \delta \pimvev}(\sigvev,0,\Svev)
        > 0  \,\, .
\ee
\end{mathletters}

{}For $\mu_B<\mu_I$, the inequalities are satisfied for $\mu_I <
m_{\pi}$, and for $\mu_B>\mu_I$, $\mu_B < m_{\pi}$.  One may check
that the second derivatives of the potential vanish on these
critical lines indicating that the transition is of second order.
Along the line $\mu_B=\mu_I>m_\pi$, the transition is first--order
since the baryon and isospin densities are discontinuous across
this line. This is exactly the phase diagram which is predicted by
chiral perturbation theory \cite{Splittorff:2001mm}.

The dispersion relations can now be calculated by examining the
quadratic terms in eq.\ (\ref{linearsig}). The Lorentz breaking
term in eq.\ (\ref{linearsig}) induces a mixing in the
$(\pi^1,\pi^2)$ and $(\pi^4,\pi^5)$ sectors. {}For the two unmixed
states, the dispersion relations are
\begin{mathletters}
\bea
\sigma \,\,\,\,\,\,\,\,\,\,\,\, E = \sqrt{{\bf p}^2+m^2+\lambda(3\sigvev^2
        +\pimvev^2 + \Svev^2)} \\
\pi^3 \,\,\,\,\,\,\,\,\,\,\,\,\,  E = \sqrt{{\bf p}^2+m^2+\lambda(\sigvev^2
        +\pimvev^2 + \Svev^2)} \ .
\eea
\end{mathletters}

The dispersion relations for the $(\pi^{1},\pi^2)$ sector
are obtained by solving
\begin{mathletters}
\bea
\det\left(\begin{array}{cc} z_1 & -2 i \mu_I E \\
        2 i \mu_I E & z_2 \end{array}\right) = 0 \,\, ,
\eea
where the diagonal terms are
\bea
z_1  \equiv {\bf p}^{\,2}-E^2+m^2+\lambda\left(\sigvev^2+
        3 \pimvev^2+\Svev^2\right)- \mu_I^2 \\
z_2 \equiv {\bf p}^{\,2}-E^2+m^2+\lambda\left(\sigvev^2+
         \pimvev^2+\Svev^2\right)- \mu_I^2 \,\, .
\eea
\end{mathletters}
The dispersion relations for the $(\pi^{4},\pi^5)$ can
be found by exchanging $\mu_I$ and $\mu_B$ in the dispersion relations
for the $(\pi^{1},\pi^2)$ sector (we denote the propagating modes in
the diquark sector by $D_\pm$).  The dispersion relations in the nonsuperfluid phase,
$\mu_I,\mu_B < m_\pi$, are
\begin{mathletters}
\bea
E_\sigma &=& \sqrt{{\bf p}^{\,2} +m^2_{\pi}+2 \lambda f_\pi^2}\\
E_{\pi_0} &=& \sqrt{{\bf p}^{\,2}+ m^2_{\pi}} \\
E_{\pi_\pm} &=& \sqrt{{\bf p}^{\,2} +  m^2_{\pi}} \pm \mu_I \\
E_{D_\pm} &=& \sqrt{{\bf p}^{\,2}+ m^2_{\pi}}\pm \mu_B \,\, .
\eea
\end{mathletters}
In the phase with the condensation of pions, the dispersion
relations are
\begin{mathletters}
\bea
E_\sigma^2 &=& {\bf p}^{\,2}+\frac{2\lambda H^2}{\mu_I^4} + \mu_I^2\\
E_{\pi_0}^2 &=& {\bf p}^{\,2} + \mu_I^2 \\
E_{\pi_\pm}^2 &=& {\bf p}^{\,2}- m^2-\frac{\lambda H^2}{\mu_I^4}+3\mu_I^2
\pm \sqrt{\left(\frac{\lambda H^2}{\mu_I^4}+m^2-3\mu_I^2\right)^2+4\mu_I^2 {\bf p}^{\,2}}\\
E_{D_\pm}^2 &=& {\bf p}^{\,2} + \mu_I^2 + \mu_B^2 \pm 2 \mu_B \sqrt{{\bf p}^{\,2} + \mu_I^2} \,\, .
\eea
\end{mathletters}
Note that for $\mu_I > m_\pi$, the $\pi_{+}$ is the massless Goldstone
boson which arises on account of the spontaneous breaking of baryon number.
In the phase with the diquark condensate, the dispersion relations can be
obtained by interchanging $\mu_I$ with $\mu_B$
and $\pi_\pm$ with $D_\pm$.  In this phase, the $D_{+}$ is the Goldstone mode.

The fermionic determinant for two-color QCD is real but not positive at
nonzero chemical potential. Therefore, it is only possible to study cases
where an equal number of flavors share a given chemical potential. The
two-flavor case discussed above is thus only accessible to lattice
simulations at either $\mu_I=0$ or $\mu_B=0$. If one doubles the number of flavors
and gives two flavors a chemical potential $\mu_u$ and the other two
flavors a chemical potential $\mu_d$, then the theory has positivity in the full
($\mu_B,\mu_I$)-phase. The phase diagram, however, has a different form than
the one above \cite{Splittorff:2001mm}.



\section{Conclusion} \label{sec:conc}

In this work, we have examined the phase structure of two-color QCD as
a function of the baryon and isospin chemical potentials as well as
the number of light quark flavors via effective Lagrangians.  We first
considered the case for which the chiral symmetry group is nonlinearly
realized.  In addition to the pseudo-Goldstone excitations, we
augmented the theory with the sector of spin-one particles and the
negative intrinsic parity terms for the group $SU(2 N_f)$.

We reviewed the salient aspects of the conformal phase transition
which is expected to occur as the number of light flavors is
increased.  Since this phase transition strongly affects the phase
structure as a function of the chemical potentials, we suggested
different ways in which lattice simulations should cast further
light on these issues.  We demonstrated that new terms in the
chemical potentials exist in the negative intrinsic parity sector
at the level of the effective Lagrangian.  Such terms are expected
to play an important role when analyzing the solitonic structure
of $N_c=2$ QCD. Since the baryons are also Goldstone bosons, the
solitonic structure is naturally richer than for $N_c=3$ QCD.
{}For example, just as $\mu_B$ couples to the winding number of
the pion sector, we showed that $\mu_I$ couples to the winding
number of the diquark sector. {}Future investigations into this
matter will certainly be interesting.

Also, unlike $N_c=3$ QCD, some of the spin-one particles of $N_c=2$
QCD are charged under baryon number.  We have hence analyzed this
sector at nonzero baryon chemical potential.  We observed novel
features such as the possible condensation of vector particles.  We
were able to calculate the value of the chemical potential at which
the vectors condense and thereby break rotational invariance.  Our
results provide new avenues for future investigations.  We then turned
to the linear effective Lagrangian including the spin-one sector.  The
phase diagram that is predicted by the linear effective Lagrangian
matches that predicted by chiral perturbation theory.

Our results can be extended to describe QCD-like theories with
quarks in the adjoint representation of the gauge group for an
arbitrary number of colors.  Another possible area of
investigation is standard QCD with a nonzero isospin chemical
potential \cite{SS}. Both cases are suitable for lattice studies.
{}For example, ordinary QCD at nonzero strange and isospin
chemical potentials has been studied in Ref.\ \cite{KT}.  The
inclusion of the vector mesons may alter the phase structure
predicted in Ref.\ \cite{KT} since the average mass splitting
between the spin-one sector and the spin-zero is not dramatically
large phenomenologically.  In fact, the spin-one sector was shown
in Ref.\ \cite{Harada:1997wr} to be crucial to the dynamics of the
light degrees of freedom.


\acknowledgments The authors thank A.\ Jackson, R.L.\ Jaffe, J.\
Schechter, J.\ Skullerud, D.\ Toublan and J.\ Verbaarschot for
useful discussions.


\appendix

\section{Explicit Realization of the $Sp(4)$ Generators}
\label{agenerators}

The generators of $SU(4)$ can be conveniently represented as
\begin{equation}
S^{a}=\left(
\begin{array}{cc}
{\bf A} & {\bf B} \\ {\bf B^{\dagger }} & -{\bf A}^{T}
\end{array}
\right) \ ,\qquad X^{i}=\left(
\begin{array}{cc}
{\bf C} & {\bf D} \\ {\bf D^{\dagger }} & {\bf C}^{T} \end{array}
\right) \ , \label{generators}
\end{equation}
where ${A}$ is hermitian, ${C}$ is hermitian and traceless,
${B}={B}^{T}$ and ${D}=-{D}^{T}$. The $\{S\}$ are also a
representation of the $Sp(4)$ generators since they obey the
relation $S^{T}E+ES=0$. We define
\begin{equation} S^{a}=\frac{1}{2\sqrt{2}}\left( \begin{array}{cc}
\tau ^{a} & {\bf 0} \\ {\bf 0} & -\tau ^{aT}
\end{array}
\right) \ ,\qquad a=1,2,3,4\ .
\end{equation}
{}For $a=1,2,3$, we have the standard Pauli matrices, while for
$a=4$ we define $\tau ^{4}={\bf 1}$. These are simply the generators for
$SU_V(2)\times U_V(1)$. {}For $a=5,\ldots,10$
\begin{equation}
S^{a}=\frac{1}{2\sqrt{2}}\left(
\begin{array}{cc}
{\bf 0} & {\bf B}^{a} \\ {\bf B}^{{a\dagger }} & {\bf 0}
\end{array}
\right) \ , \qquad a=5,\dots,10
\end{equation}
and
\begin{equation}
\begin{array}{ccc}
B^5=1 & B^7=\tau^3 & B^9 = \tau^1 \\ B^6 = i\, 1 & B^8=i\, \tau^3
& B^{10} = i \tau^1 \end{array}
\end{equation}
The five axial type generators $\{X^i\}$ are \begin{equation}
X^{i}=\frac{1}{2\sqrt{2}}\left(
\begin{array}{cc}
\tau ^{i} & {\bf 0} \\ {\bf 0} & \tau ^{iT}
\end{array}
\right) \ ,\qquad i=1,2,3\ ,
\end{equation}
where $\tau^i$ are the standard Pauli matrices. {}For $i=4,5$
\begin{equation}
X^{i}=\frac{1}{2\sqrt{2}}\left(
\begin{array}{cc}
{\bf 0} & {\bf D}^{i} \\ {\bf D}^{{i\dagger }} & {\bf 0}
\end{array}
\right) \ , \qquad i=4,5 \ ,
\end{equation}
and
\begin{equation}
D^4=\tau^2 \ , \qquad D^5=i\,\tau^2 \ .
\end{equation}
The generators are normalized as follows \begin{equation} {\rm
Tr}\left[ S^{a}S^{b}\right] ={\rm Tr}\left[ X^{a}X^{b}\right]
=\frac{1}{2}\delta ^{ab}\ ,\qquad {\rm Tr}\left[ X^{i}S^{a}\right]
=0\ .
\end{equation}

\end{document}